\documentclass[twocolumn]{jpsj2}
\usepackage{graphicx}

\title{Thermodynamics of Aggregation of Two Proteins}
\author{Kazuki \textsc{Nakanishi}$^{1,2}$
and Macoto \textsc{Kikuchi}$^{2}$
\thanks{E-mail address: kikuchi@cmc.osaka-u.ac.jp}
}

\inst{$^{1}$ Department of Physics, Osaka University, Toyonaka 560-0043\\
$^{2}$ Cybermedia Center, Osaka University, Toyonaka 560-0043\\
}

\abst{
We investigate aggregation mechanism of two proteins in a thermodynamically unambiguous manner by considering the finite size effect of free energy landscape of HP lattice protein model.
Multi-Self-Overlap-Ensemble Monte Carlo method is used for numerical calculations.
We find that a dimer can be formed spontaneously as a thermodynamically stable state when the system is small enough.
It implies the possibility that the aggregation of proteins in a cell is
triggered when they are confined in a small region by, for example, being surrounded
by other macromolecules.
We also find that the dimer exhibits a transition between unstable state and
metastable state in the infinite system.
}
\kword{protein, aggregation, prion, Monte Carlo, free energy landscape}
\begin{document}
\maketitle
\section{Introduction}
Proteins are polymers consisting of amino acids.
Under the physiological condition, each protein folds into a particular conformation called the \emph{native structure}, which determines its function.
The native structure is composed of characteristic local structures 
called $\alpha${\it -helix} and $\beta${\it -sheet}, 
both of which are stabilized by hydrogen bonds.
It has been proven experimentally by Anfinsen that the native structure is 
a thermodynamically stable state (Anfinsen's dogma).\cite{anfinsen}
The protein folding has become a subject of statistical physics from then on.
The \emph{energy landscape theory} developed in the last decade has opened a new direction of the protein folding study.
According to this theory, the landscape of the free-energy of a protein is optimized for fast folding,\cite{funnel}
and the folding is just a relaxation process in the free-energy landscape.
The theory has successfully described the folding of small globular proteins.
The free-energy landscape is now considered as a key for understanding the folding process.

Some proteins work by forming clusters {\it in vivo}, rather than as a protein monomer.
For example, hemoglobin of red blood cells forms a tetramer to transport oxygen.
On the other hand, a proteins sometimes loses its function by forming aggregates.
Moreover, fibers of aggregates called \emph{amyloid fibrils} cause some diseases.
A well-known example is the amyloid fibril of the prion proteins, which is considered to cause brain amyloidosis such as the Creutzfeldt-Jacob disease and the Bovine Spongiform Encephalopathy.
According to the widely accepted scenario, the prion protein has two different types of folded structures.
it normally takes the monomeric native structure, which is harmless.
But the very same protein happens to take an abnormally folded structure, which tends to aggregate.
Amyloid fibrils have been studied extensively by experiments and
the following properties were found: \cite{pathologic_prion,amyloid_review2}
they (1) form when the protein density is high,
(2) form when the pH is shifted from the physiological condition, and the native structure is made unstable as a result,
(3) contain the $\beta${\it -sheets}.
This third point implies that a change from the monomeric native state to the abnormal fold is accompanied by a conversion from $\alpha${\it -helix} to $\beta${\it -sheet}.

Density and temperature dependences of the stability of aggregates have been 
investigated theoretically by using lattice protein models both on 2D square lattice, \cite{effectprion} and 3D simple-cubic lattice,\cite{exploring}
and also by Protein Intermediate Resolution Model.\cite{molecular_fibril}
The following two results were obtained in common:
(1) the native structures of monomers are stable 
when both the density of the protein $\rho$ and temperature $T$ are low.
(2) the aggregates form when $\rho$ is high and $T$ is low.
Harrison {\it et al.}\cite{thermo} found further that a protein has stronger tendency of forming a homodimer when its lowest energy state is less stable.

Harrison {\it et al.}\cite{propagatable} studied the free energy landscape of a system containing two proteins by Monte Carlo simulations of 2D lattice model to discuss the stability of the aggregates.
In that study, however, two proteins were forced to contact and thus samples taken through the simulation are considered to be biased.
We should remember that a system consisting only of two proteins is thermodynamically abnormal, since it does not have a well-defined thermodynamic limit.
In fact, because of the translational entropy, a dimer cannot exist as a stable state at a finite temperature in an infinite size system.
As long as the system size is kept finite, on the other hand, thermodynamical states can be defined unambiguously.
Thus, it is essential to deal with a finite system for investigating the thermodynamics of a two protein system.
This point of view, however, has not been emphasized in the previous works, as far as we could be aware of.

In this paper, we study aggregation mechanism of two proteins 
in a thermodynamically unambiguous manner by explicitly analyzing the system size
dependence of the free-energy landscape of the HP lattice protein model.
For efficient sampling of the configurations, we use Multi-Self-Overlap-Ensemble Monte Carlo (MSOE) method proposed by
Chikenji {\it et al.};\cite{msoe_jps,msoe_aps}
in fact, this is the first application of MSOE to a multi-protein system.
The paper is organized as follows: The model and the method will be described in the next section.
Three types of the free-energy landscapes are introduced there.
Results of calculations will be presented in \S 3.
In the final section, we will discuss the thermodynamical mechanism of aggregation based on the results.

\section{Model and Method}
We use the square lattice HP model defined as follows:
(1) Each amino acid residue is represented as a bead which occupies one site of the square lattice. 
Two amino acids connected by a peptide bond are seated at adjacent sites; namely, the length of the peptide bond is taken as a lattice constant.
Then a conformation of a protein consisting of $N$-amino acids is represented as
 a self-avoiding walk of length $N$.
(2) Each amino acid is either of two types, H(Hydrophobic) or P(Polar).
(3) Only two non-bonded H beads that are nearest neighbors have energy equal to -1. 
Such a pair of H beads is said to make an HH-\emph{contact}. 
Other contacts, HP and PP, do not contribute to energy.
Then the contact energy is given as follows:
\begin{eqnarray}
 E &=& \sum_{i<j+1} u(S_i,S_j) \Delta ({r_i,r_j}) , \label{energy} 
\end{eqnarray}
where $r_i$ represents the position of the $i$-th residue.
$\Delta (r_i,r_j)$ is equal to 1 if $r_i$ and $r_j$ are the nearest neighbors. 
Otherwise, $\Delta (r_i,r_j)$ is equal to 0.
$S_i$ stands for H or P.
$u(S_i,S_j)$ represents the interaction energy
between the $i$-th amino acid and the $j$-th one:
\begin{equation}
u({\rm H,H})=-1,u({\rm H,P})=u({\rm P,H})=u({\rm P,P})=0. \label{u_prop}
\end{equation}
This definition of the energy takes only the effect of hydrophobicity into account.
It should be noted that the interaction with the solvent is included in the contact energy effectively, and thus the contact \emph{energy} actually is the excess free energy of forming an HH-contact.

\begin{figure}
  \includegraphics[width=3.5cm]{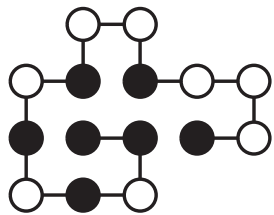}
  \caption{The lowest energy structure of the monomer of the
    sequence 1 ($E=-6$). 
	Black and white beads represent H and P
residues, respectively.}
  \label{fig1}
  \end{figure}

\begin{figure}
  \includegraphics[width=7cm]{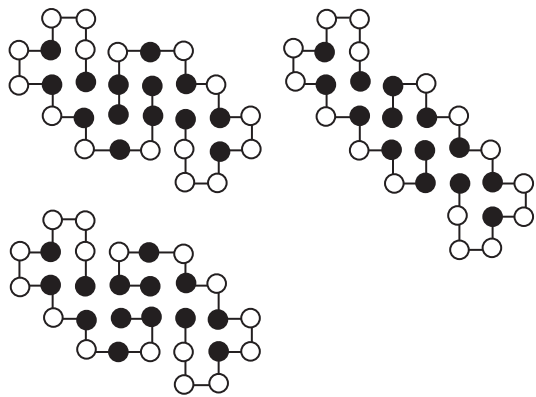}
  \caption{All the lowest energy structures of the dimers of the
    sequence 1 ($E=-14$).
	Black and white beads represent H and P residues, 
	respectively.}
  \label{fig2}
\end{figure}

\begin{figure}
  \includegraphics[width=3.5cm]{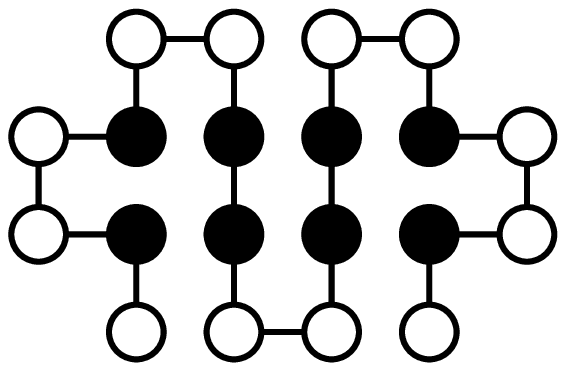}
  \caption{The lowest energy structure of the monomer of the
    sequence 2 ($E=-8$). 
	Black and white beads represent H and P residues, respectively.}
    \label{fig3}
 \end{figure}

\begin{figure}
  \includegraphics[width=7cm]{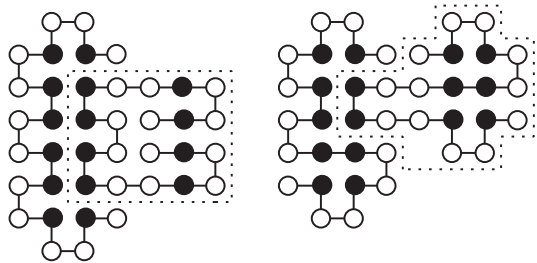}
  \caption{Examples of the lowest energy structures of the dimer of the sequence 2
  ($E=-18$). 
  Black and white beads represent H and P residues, respectively.}
  \label{fig4}
\end{figure}

Two HP sequences will be studied in the following sections.
The sequence 1 is ${\rm H^2 PHPHPHP^2 H P^3 H}$ (15 residues), which was used by Harrison {\it et al.}.\cite{thermo}
The monomer of this sequence has the nondegenerate lowest energy structure ($E=-6$) shown in Fig. \ref{fig1}.
The same interaction energy, eq.(\ref{u_prop}), is applied also to
the inter-chain contacts when two chains are considered.
Then the lowest energy of two chains is that of a dimer ($E=-14$), which is three-fold degenerate as shown in Fig. \ref{fig2}.
The sequence 2 is ${\rm PHP^2HP^2H^2P^2H^2P^2HP^2HP}$ (20 residues) used by Gupta {\it et al.}.\cite{effectprion}
Figure \ref{fig3} shows the lowest energy structure ($E=-8$) of the monomer
of this sequence, which also is nondegenerate. 
But the lowest energy of a dimer is equal to that of two monomers,
if eq.(\ref{u_prop}) is applied to the inter-chain contacts. 
In order to make the lowest energy structures of 
a dimer more stable than two independent monomers, 
we change $u$(H,H) to $-1.5$ only for the \emph{inter-chain} HH-contacts.
As a result of this modification, the lowest energy becomes $E=-18$.
There are many conformations of the lowest energy, some of which are shown in Fig. \ref{fig4}.
The lowest energy dimer states are not simply composed of two native states of monomers for neither of two sequences.
In other words, the monomers are required to unfold partially before forming a dimer.

We introduce the following three types of free energy, all of which characterize the aggregation process:
\begin{enumerate}
\item 
The free energy as a function of 
the distance $R$ between the centers of mass of two chains:
\begin{eqnarray}
F(R)&=&-T \left. \log  \int e^{-\frac{E}{T}} d E \right|_{R-\Delta r \le r<R+\Delta r},  \label{freer} 
\end{eqnarray}
This function is expected to behave asymptotically for large $R$ as $F(R) \sim -T\log R$,
because it is composed by the translational entropy and the free energy of monomers when two chains are far apart and cannot interact.
\item
The free energy as a function of $R$ and the number of inter-chain HH-contacts $N_{inter}$.
Since $N_{inter} \ne 0$ indicates that a dimer is formed, we call this free energy $F_{di}$, where the suffix stands for \emph{dimer}:
\begin{equation}
F_{di} (N_{inter},R)=- T \left. \log \int 
 e^{-\frac{E}{T}} d E \right|_{R -\Delta r\le r<R+\Delta r,N_{inter}={\rm fixed}} . \
 \label{fcon}
\end{equation}
The bin size $\Delta r = 0.5$ is used throughout the present work.
\item
The free energy as a function of $R$ and the number of intra-chain HH-contacts $N_{intra}$ calculated only for configurations of $N_{inter}=0$, namely two independent monomers.  Then we call this free energy $F_{mono}$.
\begin{equation}
F_{mono}(N_{intra},R) 
 = - T \left. \log \int
 e^{-\frac{E}{T}} d E \right|_{R -\Delta r \le r<R+\Delta r,N_{intra}={\rm fixed},N_{inter}=0}  .\label{fcon2}
\end{equation}
\end{enumerate}
In the actual calculations, summation over Monte Carlo samples are taken instead of the integration. 

It is difficult to obtain the lowest energy conformations by Monte Carlo
simulations because of the topological barrier due to the excluded volume effect.
In order to overcome this difficulty, we employ MSOE, in which the excluded volume condition is systematically weakened.
For that purpose, we define an effective energy associated with overlap as follows:
$V=\sum_{i} (M_i-1)^2$ for $M_i \geq 1$ and $V=0$ for $M_i=0$, 
where $M_i$ is the number of amino acids
on the $i$-th lattice site.
The procedure of MSOE for two chains is the following:
\begin{enumerate}
\item
$M_i$ is counted without distinction between two chains.
\item
Bivariate multicanonical Monte Carlo method is applied 
to obtain a flat distribution for both $V$ and $E$.
\item
By taking only non-overlapping conformations, that is, samples with $V=0$,
the standard multicanonical ensemble\cite{multicanonical} is obtained. 
\item
Free energy and other thermodynamic quantities are calculated through the histogram reweighting method.
\end{enumerate}

We treat systems of different sizes.
For the sequence 1, the edge lengths $L$ of the systems 
are 60, 80, and 100 (the unit of the length is the lattice constant).
For the sequence 2, $L=60$, 80, 100, and 120.
Periodic boundary conditions are imposed.

\section{Result}
\subsection{Sequence 1}
\begin{figure}
    \includegraphics[width=7cm]{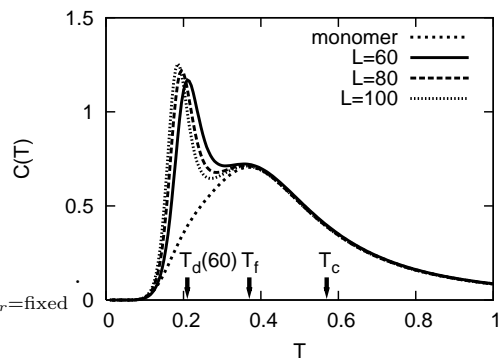}
    \caption{     
	  Specific heat per residue of two chains for $L=60, 80,$ and $100$,
	  and that of monomer  of the sequence 1.
      $T_f$ is the folding temperature of the monomer,  $T_d(L)$ is the size-dependent dimerization temperature.
      The unstable-metastable transition temperature, described in  \S 4, is indicated by $T_c$.
    }
    \label{fig5}
   \end{figure}

Figure \ref{fig5} shows the specific heat per residue $C(T)$ of two chains for $L=60, 80,$ and $100$, and that of the monomer of the sequence 1.
$C(T)$ of the monomer exhibits a single peak.
The monomer folds into the native state for lower temperature than the peak.
Thus we call this temperature of the peak the folding temperature $T_f$.
On the other hand, two peaks are seen in $C(T)$ of the two chains;
one is at $T_f$ and additional one is at a lower temperature.
As we will see later from $F_{di}$, two chains change into a dimer at the additional peak.
This peak shifts to lower temperature as $L$ increases.
We call the temperature of this peak the dimerization temperature $T_d(L)$, below which the dimer state is thermodynamically stable.

\begin{figure} 
    \includegraphics[width=7cm]{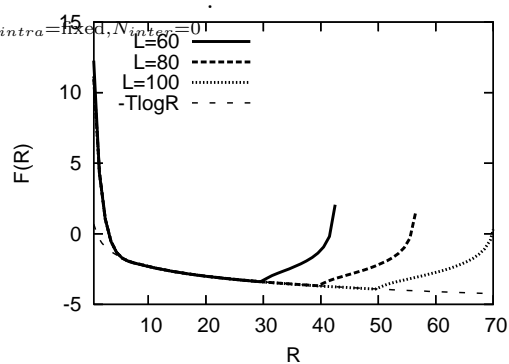}
     \caption{$F(R)$ of the sequence 1 at $T=1.0$ for $L=60, 80$, and 100.
	$-T \log R$ is also shown.
}
     \label{fig6}
\end{figure}

\begin{figure}
    \includegraphics[width=7cm]{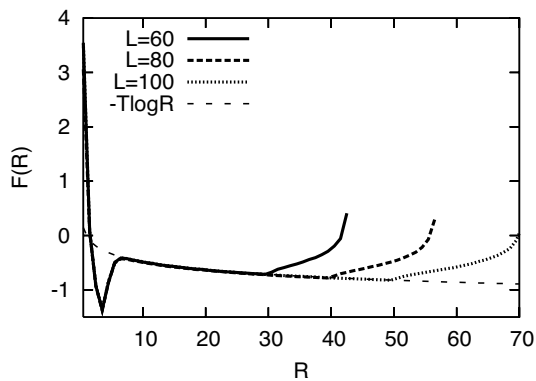}
    \caption{
      $F(R)$ of the sequence 1 at $T_d(60)=0.21$ for $L=60, 80$, and 100.
     $-T \log R$ is also shown.
		  }
    \label{fig7}
    \end{figure}

Figures \ref{fig6} and \ref{fig7} show $F(R)$ at high temperature ($T=1.0 > T_f$) and at $T_d(60)$, respectively.
Since $F(R)$ is calculated up to an arbitrary additive constant, 
we adjusted it so that $F(R)$ coincides with $-T \log R$ for large $R$.
$F(R)$ at high temperature shows the following features:
(1) $F(R) \sim -T\log R$ for larger $R$ than about $10$.
It is rather a trivial consequence of the translational entropy and indicates that Monte Carlo samples are taken properly.
(2) it turns to increasing for $R>L/2$, because of the reduction of the translational entropy due to the finite size effect.
As a result, $F(R)$ has a single minimum at $R=L/2$.
$F(R)$ at $T_d(60)$, on the other hand, exhibits an additional minimum which corresponds to the dimer state.
In other words, the independent monomer state and the dimer state are separated by a free-energy barrier.
$F(R)$ near the dimer state exhibits no evident system size dependence.
The dimer state thus can either be stable or metastable, depending on the system size.

\begin{figure} 
      \includegraphics[width=7cm]{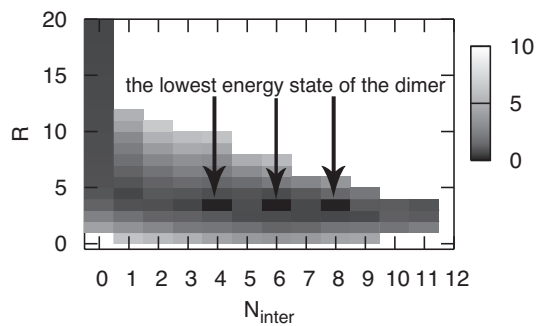}
      \caption{$F_{di}(N_{inter},R)$ of the sequence 1 $(T=T_d(60), L=60)$. Arrows indicate the locations of three lowest energy dimer states.
	   }
      \label{fig8}
    \end{figure}

\begin{figure}
      \includegraphics[width=7cm]{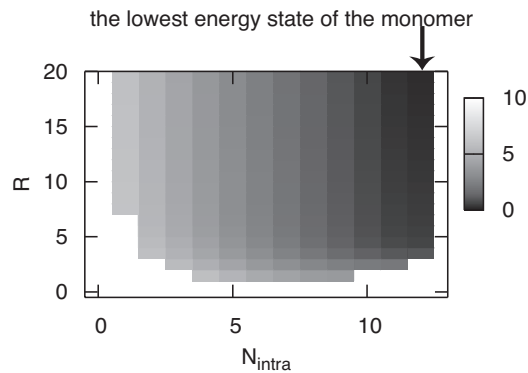}
      \caption{$F_{mono}(N_{intra},R)$ of the sequence 1 $(T=T_d(60),L=60)$.
	  }
      \label{fig9}
\end{figure}

Dimerization process can be deduced from $F_{di}$ and $F_{mono}$
shown in Figs. \ref{fig8} and \ref{fig9}, respectively, for $L=60$ at $T=T_d(60)$.
$F_{di}$ shows that the behavior of two chains changes at $R = R_c \sim 5$.
The inter-chain HH-contact is hardly formed for $R > R_c$, while a dimer ($N_{inter}\ne 0$) is more stable than two monomers ($N_{inter}=0$) for $R < R_c$.
Three lowest energy dimer configurations are in the minima of the free energy indicated by the arrows.
The independent monomer state and the dimer state are separated by a free-energy barrier at $(N_{inter},R) \simeq (1,R_c)$.
Thus one (or at most two) inter-chain contact is formed at the transition state.
$F_{mono}$ suggests that two chains are mostly in the native state individually for $R > R_c$, while the native state becomes unstable rapidly as $R$ decreases for $R<R_c$.
Considering together these two figures and the fact that the lowest energy state of the monomers should be unfolded partially to become a dimer,
we can infer that the dimerizaion takes place through the following processes:
(1) Two individually folded chains can approach each other as close as $R \sim R_c$.
(2) Once one inter-chain contact or two is formed through partial unfolding by thermal fluctuation, the chains fold into a dimer rapidly. 

\subsection{Sequence 2}

\begin{figure} 
    \includegraphics[width=7cm]{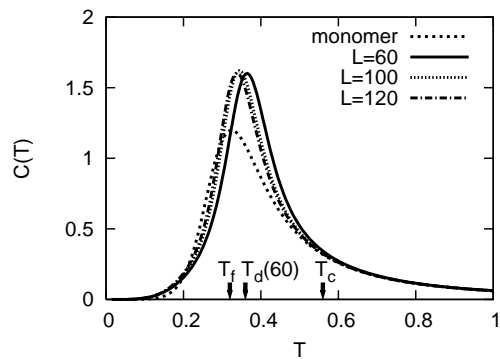}
     \caption{Specific heat per residue of two chains for $L=60, 80, 100$, and 120,
      and that of monomer of the sequence 2.
      $T_f$ is the folding temperature of the monomer,  $T_d(L)$ is the size-dependent dimerization temperature.
      The unstable-metastable transition temperature, described in  \S 4, is indicated by $T_c$.}
   \label{fig10}
   \end{figure}

Figure \ref{fig10} shows $C(T)$ of two chains for $L=60, 80, 100,$ and $120$, and that
of monomer of the sequence 2.
In contrast to the sequence 1, only one peak is seen for $C(T)$ of the two chains, which is located at higher temperature than $T_f$.
This peak shifts to lower temperature as $L$ increases.
As we will see later in $F_{di}$, two chains fold into a dimer near the peak.
Thus the temperature of this peak corresponds to $T_d(L)$.
So, in this case the dimerization takes place at higher temperature than $T_f$ as a result of the larger inter-chain HH-contact energy;
the monomeric native state never realizes as a thermodynamically stable state.

\begin{figure} 
    \includegraphics[width=7cm]{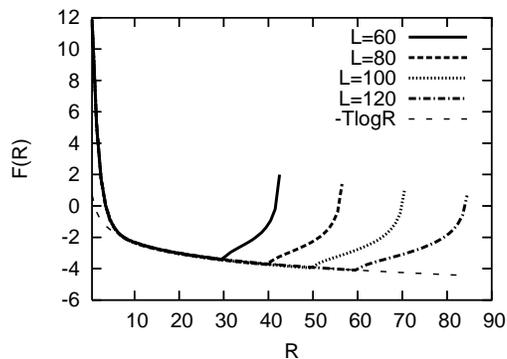}
    \caption{
      $F(R)$ of the sequence 2 at $T=1.0$ for $L=60, 80, 100,$ and 120.
    $-T \log R$ is also shown.
    }
    \label{fig11}
  \end{figure}

\begin{figure} 
      \includegraphics[width=7cm]{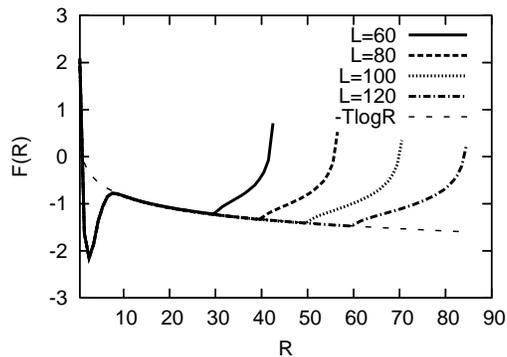}
      \caption{
        $F(R)$ of the sequence 2 at $T_d(60)=0.36$ for $L=60, 80, 100$, and 120.
        $-T \log R$ is also shown.  
      }
      \label{fig12}
    \end{figure}

\begin{figure} 
      \includegraphics[width=7cm]{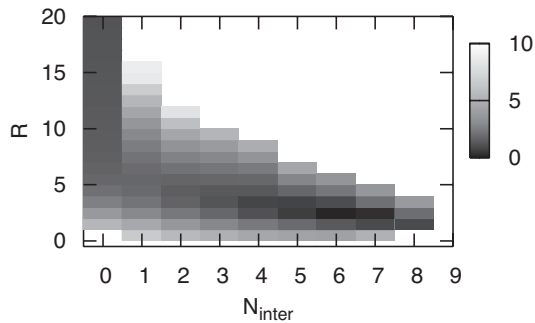}
      \caption{$F_{di}(N_{inter},R)$ of the sequence 2 $(T=T_d(60), L=60)$.
	   }
      \label{fig13}
    \end{figure}

\begin{figure}
      \includegraphics[width=7cm]{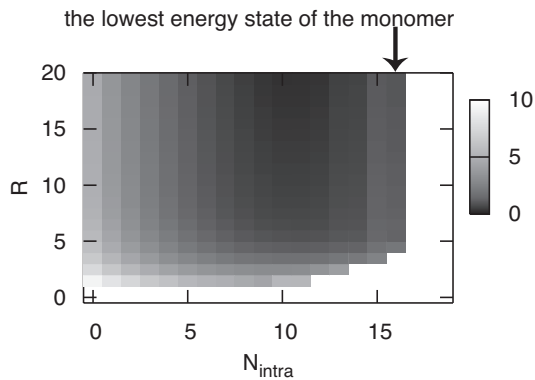}
      \caption{$F_{mono}(N_{intra},R)$ of the sequence 2 $(T=T_d(60),L=60)$.
	  }
      \label{fig14}
\end{figure}

Figures \ref{fig11} and \ref{fig12} show $F(R)$ at high temperature ($T=1.0 > T_f$) and at $T_d(60)$, respectively.
Qualitative features are similar to those of sequence 1.
So, a dimer also for this sequence is either a stable state or a metastable state  for $T<T_d(L)$, depending on the system size.
$F_{di}$ and $F_{mono}$ for for $L=60$ at $T=T_d(60)$ are shown in Figs. \ref{fig13} and \ref{fig14}, respectively.
Qualitative features are again similar to those of sequence 1.
The largest difference is seen in $F_{mono}$ that the individual monomers are partially unfolded for this sequence, which is a natural consequence of $T_d(60) > T_f$.
In contrast to the sequence 1, the lowest energy dimer states consist a broad ensemble, which distribute around the free-energy minimum.
The transition state between the individual monomer state and the dimer state is again located at $(N_{inter},R) \simeq (1,5)$.
A possible scenario of forming a dimer is also similar as that of sequence 1, except that the individual monomers are partially unfolded even before making an inter-chain contact.

\section{Summary and Discussion}
We have investigated the free energy landscape of two protein systems in a
thermodynamically unambiguous way by considering the finite size effect properly.
We found that a dimer is thermodynamically stable for $T<T_d(L)$, and
$T_d(L)$ decreases as $L$ becomes larger.

On the basis of the present results, we can discuss the stability of a dimer
in, for example, a confined system such as two proteins surrounded by
other macromolecules in a cell, although the periodic boundary conditions were
actually used in the present work.
Then the following scenario for the early stage of aggregation is suggested:
if two proteins are confined in a sufficiently small region,
a dimer is formed spontaneously because it is \emph{thermodynamically stable} rather than metastable;
once a dimer is formed, it will survive for a while as a metastable state even after released from the confined region, because of the free energy barrier.
It should be noted that although this scenario seems to be consistent with
results of experiments and other theoretical studies on aggregation, what we have discussed above is not just a condition for dimerization to take place such as high density, but a plausible mechanism of spontaneous dimerization based on thermodynamic stability.

Now we readily consider the limit of $L \to \infty$.
As we have already seen, $F(R)$ shows the asymptotic $-\log R$ dependence as long as $R < L/2$, while $F(R)$ near the dimer state hardly depends on the system size.
Thus the minimum free energy of the independent monomer state is proportional to $-\log L$ and can become infinitely low in this limit.
The dimerization temperature $T_d(L)$, which is determined by the free-energy balance between these two states, is expected to vanish at the same time.
The folding temperature $T_f$, on the other hand, does not depend on the system size and stays finite.
Accordingly, as the system size increases, $T_d(L)$ should eventually becomes lower than $T_f$ irrespective of the sequence.
As a result, the single peak observed in $C(T)$ for the sequence 2, which corresponds to the dimerization, will split into two as the sequence 1, if the system becomes sufficiently large so that the folding of monomer takes place at higher temperature than dimerization.
Although the dimer can never exist as a thermodynamically stable state in the infinite size limit, stability of the dimer actually changes with temperature;
The dimer is unstable at high temperature, while it becomes metastable at low temperature.
This change is not a phase transition in a strict sense, 
but we may call it \emph{unstable-metastable transition} of the dimer.
The temperature at which this transition takes place is indicated as $T_c$ in 
Figs. \ref{fig5} and \ref{fig10}.
As seen in the figures, no significant effect is observed  in $C(T)$ at $T_c$.
This temperature, however, still has a well-defined physical meaning that the dimer never becomes stable above it even in a finite system.

Recently, Levy et al. \cite{binding} have studied formation of a homodimer for several different proteins based on the free energy landscape calculation.
They used an off-lattice model with Go-like interactions both for intra- and inter-chain interactions.
Thus the ground state of the dimer is unique and the same intra-chain contacts are formed in the monomeric native state and the dimer ground state.
System size dependence was not explicitly considered as the previous works described in the introduction did.
Instead, two chains are made unseparable by additional virtual forces.
Therefore, their model and the computational framework are totally different from the present work.
But it will still be informative to compare the result with the present one.
They found that the structure of the free-energy landscape differ largely for different proteins reflecting a variety of binding mechanisms.
Among them, the free-energy landscape of $\lambda$ repressor shares a common feature with that of sequence 1.
In fact, according to their result, two independently folded $\lambda$ repressor change into a dimer when they come close enough.
This protein is known to form a dimer through so called \emph{induced fit} mechanism.
So, we may say that the sequence 1 studied in the present work also exhibits the induced fit into one of the three ground states.
The folding process of sequence 2 differs from that of sequence 1 in that the monomers are partially unfolded before forming a dimer.
It seems to correspond to \emph{folding-binding reaction} of, for example, Arc-repressor.
We should stress, however, that the folding process is expected to change into the induced-fit type as sequence 1, if the system becomes large enough so that the dimerization temperature is lower than the folding temperature.

Resemblances and differences with the crowding effect of protein folding is worth mentioning.
It has been pointed out by Takagi {\it et al.}\cite{crowding} that free energy landscape of a protein is modified when it is confined in a small region, and a faster folding is achieved as a result.
The present study may be regarded as its dimer counterpart.
In contrast to that study, however, we found that the metastable state of a dimer can change to the stable state as a consequence of confinement.

As a final remark, we stress again that the finite-size effect is a key for understanding the aggregation of proteins.
The method proposed in this paper can be applied straightforwardly to the aggregation of more realistic protein models and that of three or more proteins.

\section*{Acknowledgment}
We thank K. Tokita and F. Takagi for fruitful discussion.
The present work is partially supported by IT-program of Ministry of Education,
Culture, Sports, Science and Technology, The 21st Century COE program named
"Towards a new basic science: depth and synthesis" and Grant-in-Aid for Scientific
Research (C) (17540383) from Japan Society for the Promotion of Science.

\end{document}